\documentclass[12pt]{iopart}

\newcommand{\be}{\begin{equation}}
\newcommand{\ee}{\end{equation}}
\newcommand{\ben}{\begin{eqnarray}}
\newcommand{\een}{\end{eqnarray}}
\newcommand{\n}{\label}
\newcommand{\no}{\noindent}

\newcommand{\La}{\Lambda}
\newcommand{\ga}{\gamma}

\newcommand{\ba}{\begin{eqnarray}}
\newcommand{\ea}{\end{eqnarray}}
\newcommand{\ov}{\overline}

\newcommand{\al}{\alpha}

\usepackage{color}
\begin{document}
\title[Phantom
cosmologies and fermions]{Phantom
cosmologies and fermions}

\author{Luis P. Chimento$^1$, Fernando P. Devecchi$^2$,
M\'onica Forte$^1$ and Gilberto M. Kremer$^2$}

\address{$^1$ Departamento de F\'{\i}sica, Facultad de Ciencias Exactas y
Naturales,  Universidad de Buenos Aires, Ciudad
Universitaria, Pabell\'on I, 1428 Buenos Aires, Argentina}
\address{$^2$ Departamento de F\'\i sica,
Universidade Federal do Paran\'a, Caixa Postal 19044, 81531-990 Curitiba, Brazil}

\ead{chimento@df.uba.ar, devecchi@fisica.ufpr.br, monicaforte@fibertel.com.ar,  kremer@fisica.ufpr.br}

\begin{abstract}
Form invariance transformations can
be used for constructing phantom cosmologies starting with conventional cosmological models. In this work we reconsider the scalar field case and extend the discussion  to fermionic fields, where the ``phantomization" process exhibits a new class of possible accelerated regimes. {As an application we  analyze the cosmological constant group for a fermionic seed fluid.}
\end{abstract}

\submitto{\CQG}
\maketitle

%%%%%%%%%%%%%%%%%%%%%%%%%%%%%%%%%%%%%%%%%%%%%%%%%%%%%%%%%%%%%%%%%%%
\section{Introduction}
%%%%%%%%%%%%%%%%%%%%%%%%%%%%%%%%%%%%%%%%%%%%%%%%%%%%%%%%%%%%%%%%%%%

The issue of form invariance symmetries \cite{Chimento:2002gb}
 has been invoked recently to extract different evolution regimes
 from Friedmann-Robertson-Walker (FRW) and Bianchi type V cosmologies \cite{Aguirregabiria:2003uh}.
This is the case when we obtain the so-called phantom cosmologies
\cite{Chimento:2003qy}-\cite{Chimento:2005wv}
from ordinary regimes like universes filled with barotropic fluids \cite{Caldwell:1999ew}.
These phantom models generate solutions that contain negative pressure situations, therefore promoting positive accelerated
expansions \cite{Johri:2003rh},\cite{Dabrowski:2003jm}.
 The  case where the gravitational source is a bosonic field was
investigated in \cite{Chimento:2002gb},\cite{Chimento:2003qy},\cite{Chimento:2004br}, and the question that follows would be what happens when the field promoting the  universe's expansion is a fermion whose dynamics is described by the Dirac equations. This kind of model has been analyzed in several works \cite{Obukhov:1993fd}, \cite{Ribas:2005vr} where it was shown that regimes with  ordinary matter/dark energy transitions  are possible solutions of the
Einstein and Dirac equations. Taking these ideas into account, the main purpose of this work is to investigate the form invariance transformations (FIT) that result when the Klein-Gordon dynamics is replaced by the
Dirac equations in a FRW space-time.
  The paper is structured as follows:  in section II we review the basic ideas behind the FIT in cosmology; in section III we focus on the scalar field case. In section  IV we analyze the symmetries behind the fermionic formulations {and analyze the cosmological constant group with a fermionic seed fluid.} In section V we introduce the dual transformation and the associated "phantomization" process. Finally in section VI we present our conclusions.

%%%%%%%%%%%%%%%%%%%%%%%%%%%%%%%%%%%%%%%%%%%%%%%%%%%%%%%%
\section{Internal symmetry in FRW}
%%%%%%%%%%%%%%%%%%%%%%%%%%%%%%%%%%%%%%%%%%%%%%%%%%%%%%%

    An interesting method of obtaining  new exact solutions of the Einstein equations from already existing ones, by using the  FIT, has been developed in a series of papers \cite{Chimento:2002gb}-\cite{Chimento:2005wv},  \cite{Cataldo:2005gb}-{\cite{Chimento:2006gk}.

In spatially flat perfect fluid FRW cosmologies, such transformations can be viewed as a prescription relating the quantities $a$, $H$, $\rho$ and $p$ in a given initial scenario to quantities $\bar  a$, $ \bar H$, $\bar \rho$ and $\bar p$ corresponding to a new cosmological model.

 As a starting point we present a quick review of this internal symmetry with the purpose of investigating it in bosonic and fermionic cases. As is well-known, the Einstein
equations for a flat FRW cosmological model with scale factor $a$, and filled with a perfect fluid with energy density $\rho$ and pressure $p$, are
\ba \n{00} &&3H^2=\rho,\\
\n{con} &&\dot \rho+3H(\rho+p)=0,
\ea

\noindent where $H=\dot a/a$.
The corresponding FIT is given
by \cite{Chimento:2002gb}
\ba\n{11}
 &&\bar\rho=\bar\rho(\rho)\;,\n{tr1}\\
 &&\bar H=\left(\frac{\bar
\rho}{\rho}\right)^{{1}/{2}}H\;,\label{tr2}\\ &&\bar p=-\bar
\rho+\left(\frac{\rho}{\bar\rho}\right)^{1/2}(\rho+p)\frac{d\bar
\rho}{d\rho}\; ,\label{tr3} \ea
 where $\bar\rho=\bar\rho(\rho)$ is an
arbitrary invertible function and the set of transformations (\ref{tr1})-(\ref{tr3}) give rise to the form invariance symmetry (FIS) group . We are interested in analyzing the perfect fluid case, where heat conduction, shear and bulk viscosities are absent. Furthermore, we assume a  barotropic equation of state for the pressure which reads
 $p=(\gamma-1)\rho$. Hence, it follows that barotropic indices of the original $\gamma$ and transformed fluid $\bar\gamma$ are related by
 \be \n{tg} \bar \gamma= \left
(\frac{\rho}{\bar\rho}\right)^{3/2}\frac{d\bar \rho}{d\rho}\,\gamma . \ee

\noindent For later application it will be useful to investigate the transformation rule for the quantities which characterize the geometry and the fluid when the transformation is generated by \cite{Chimento:2002gb}
\ben
\n{tr}
\bar \rho = n^2\rho,
\een
where $n$ is the constant parameter of the FIS group. Hence, for the geometrical quantities we get
\ben
\n{tha}
\bar H=nH,   \quad\to\quad \bar a=a^n,
\een
and
\ben\n{12}
\bar w=\bar\ga-1=\frac{\ga}{n}-1,\n{bw}\\
\bar p=\bar w\bar\rho=\left(\frac{\ga}{n}-1\right)\bar\rho,\n{bp}
\een
for the state parameter $w$ and the pressure of the fluid.

Let us analyze the equation of state (\ref{bp}) generated by transformation (\ref{tr}) when the seed fluid represents a normal fluid with $\ga>0$. We can infer: (i) when $0<n<\ga$ it follows that $\bar w>0$ and the cosmological fluid behaves as a normal fluid; (ii) when $n>\ga$ it follows that $-1<\bar w<0$ and the cosmological fluid behaves as a quintessence constituent; (iii) when $n<0$ it follows that $\bar w<-1$  and the cosmological fluid behaves as a phantom constituent.

%%%%%%%%%%%%%%%%%%%%%%%%%%%%%%%%%%%%%%%%%%%%%%%%%%%%%%%%%%%%%%%
\section{The bosonic case}
%%%%%%%%%%%%%%%%%%%%%%%%%%%%%%%%%%%%%%%%%%%%%%%%%%%%%%%%%%%%%%%%

\noindent In this section we analyze the transformation rules for a scalar field $\phi$, self-interacting through a potential $V(\phi)$, under the simple transformation generated by (\ref{tr}) \cite{Chimento:2002gb}.
Applying Eqs. (\ref{tr2}) and (\ref{tr3}) to the dynamical quantities associated with the scalar field, i.e., its energy density $\rho_\phi=\dot\phi^2/2+V(\phi)$ and pressure  $p_\phi=\dot\phi^2/2-V(\phi)$, we obtain
\ba
\n{tp}
&&{\dot {\bar\phi}}\,^2=n\dot {\phi}^2,\\ &&
\n{tv}
\bar V(\bar
\phi)=n(n-1)\frac{{\dot\phi}^2}{2}+n^2 V(\phi).
\ea
Note that $\bar V(\bar\phi)$  given in the above equation is not properly a potential, since it depends on the kinetic and potential terms of the seed fluid, but it will become a potential once the cosmological solution of the seed fluid is known. In other words, the FIT acts on the internal space of the Friedmann's cosmology so the cosmological time is not affected by those kinds of transformations. The knowledge of the seed solution $\phi(t)$ allows us to invert it obtaining $t=t(\phi)$, $\dot\phi(t)=\dot\phi(t(\phi))$ and $\dot\phi=\dot\phi(\phi)$. Using the latter in Eq. (\ref{tp}) we get $\dot{\bar\phi}^2=n\,\dot\phi^2(\phi)$ and $\bar\phi=\sqrt{n}\,\phi+\phi_0$ where $\phi_0$ is an integration constant. Then, the kinetic and the potential terms become $\dot\phi^2(\phi)=\dot\phi^2((\bar\phi-\phi_0)/\sqrt{n})$ and $V(\phi)=V((\bar\phi-\phi_0)/\sqrt{n})$, respectively. Inserting these last two relations into the FIT (\ref{tp}) and (\ref{tv}) leads to the transformed potential $\bar V(\bar\phi)$ which is expressed as a function of the transformed potential $\bar\phi$.

In order to study the action of the group,  we will now generate the set of power law solutions starting from a particular seed one.

%%%%%%%%%%%%%%%%%%%%%%%%%%%%%%%%%%%%%%%%%%%%%%%%%%%%%%%
\subsection{Bosonic power law group}
%%%%%%%%%%%%%%%%%%%%%%%%%%%%%%%%%%%%%%%%%%%%%%%%%%%%%%%%%%%

Let us solve the Einstein-Klein-Gordon equations for a free scalar field, $V=0$. The solutions of those equations are given by
\be
\n{seed1}
a^{(\pm)}=(\pm\, t)^{1/3}, \quad \phi=\pm\sqrt{\frac{2}{3}}\ln{|t|}, \quad V=0.
\ee

\no where the branches $a^{(+)}$ and $a^{(-)}$ are defined for $t>0$ and $t<0$ respectively. Here we call attention  to the fact that the $\pm$ branches of the scalar field $\phi$
are not related to the branches $a^{(+)}$ and $a^{(-)}$ of the cosmic scale factor. Now, we use the latter as a seed set for the FIS group. To do that, we first calculate the transformed quantities when the group is acting on the above solutions
\ba
\n{tp1}
&&{\dot {\bar\phi}}\,^2=n\dot {\phi}^2,\\ &&
\n{tv1}
\bar V(\bar
\phi)=n(n-1)\frac{{\dot\phi}^2}{2}.
\ea

\no Finally combining the seed set (\ref{seed1}) with the above equations we obtain the power law solutions together with the scalar fields and corresponding potentials \cite{Chimento:2004br}
\ba
\n{ta11}
&&{\bar a^{(\pm)}=(\pm\, t)^{n/3}},\\
\n{tp11}
&&{\bar\phi}=\pm\sqrt{\frac{2n}{3}}\ln{|t|},\\ &&
\n{tv11}
\bar V(\bar
\phi)=\frac{n}{3}(n-1)e^{\mp\sqrt{6/n}\,\bar\phi}.
\ea

\no For the identity transformation, $n=1$, the latter equations (\ref{ta11})-(\ref{tv11}) reduce to the seed solution (\ref{seed1}). We conclude that power law solutions can be generated from a seed solution corresponding to the free scalar field.

%%%%%%%%%%%%%%%%%%%%%%%%%%%%%%%%%%%%%%%%%%%%%%%%%%%%%%%%%%%%%%%%%%%%%%%%%
\section{The fermionic case}
%%%%%%%%%%%%%%%%%%%%%%%%%%%%%%%%%%%%%%%%%%%%%%%%%%%%%%%%%%%%%%%%%%%%%%%%

\noindent In this section we extend the FIT to a fermionic field satisfying the Dirac equation in curved space time. As required by equations (1) and (2) we must  calculate the energy density and the hydrostatic pressure of the fermionic field. We have \cite{Ribas:2005vr}
\be
\n{r}
\rho_\psi=m(\ov{\psi} \psi)+V,
\ee
\be
\n{p}
p_\psi= \frac{dV}{d\psi} \frac{\psi}{2}+\frac{\ov {\psi}}{2}
\frac{dV}{d\ov {\psi}} - V,
\ee
where  $\psi$ and $\overline{\psi}=\psi^\dag\ga^0$ are
 the spinor field and its adjoint, respectively.

To obtain the transformation properties of $\rho_\psi$ and $p_\psi$ we express the potential in terms of the scalar invariant $X = (\ov {\psi} \psi)^2$
 and the pseudo-scalar invariant $Y = (i\ov {\psi }\gamma ^5 \psi)^2$. To this end, we assume a generic potential $V=V(X,Y)$, that includes, among others, the Nambu-Jona-Lasinio potential \cite{Ribas:2005vr}. For this kind of potential the  hydrostatic pressure becomes

\be
\n{pxy}
p_\psi = 2X\frac{\partial V}{\partial X}+2Y\frac{\partial V}{\partial Y}-V. \ee

Applying the transformation rules (\ref{tr1})-(\ref{tr3}), (\ref{tr}) to the energy density (\ref{r}) and the hydrostatic pressure (\ref{pxy}) we find that
\be
\n{trf}
\bar m \sqrt{\bar X}+\bar V=n^2(m\sqrt{X}+V),
\ee
\be\n{tpf}
2\bar X\frac{\partial\bar V}{\partial\bar X}+2\bar Y\frac{\partial\bar V}{\partial\bar Y}+\bar m\sqrt{\bar X}
=
n\left[2X\frac{\partial V}{\partial X}+2Y\frac{\partial V}{\partial Y}+m\sqrt{X}\right].
\ee

\no Inserting $\bar V$, from Eq.(\ref{trf}), into Eq.(\ref{tpf}) we obtain the following equation
$$
\left({m\over2\sqrt{X}}+\frac{\partial V}{\partial X}\right)
\left[\bar X \frac{\partial X}{\partial \bar X} + \bar Y
\frac{ \partial X}{\partial \bar Y}-{X\over n}\right]
$$
\be
\n{def}
+\frac{\partial V}{\partial Y}
\left[\bar X \frac{\partial Y}{\partial \bar X} + \bar Y
\frac{ \partial Y}{\partial \bar Y}-{Y\over n}\right]=0.
\ee

\no All transformations satisfying the last condition (\ref{def}) represent internal symmetries of the Einstein equations with a fermionic source, provided that condition becomes an identity for any potential and particle mass. In any other case, the above equation becomes a partial differential equation whose solutions would lead to particular  transformations depending on the transformed quantities. We must discard them because a symmetry transformation depends on a set of parameters or functions which do not include the quantities to be transformed. Hence, a necessary and sufficient condition for Eq. (\ref{def}) to be satisfied for any potential and particle mass is that the two terms in square brackets must vanish separately, i.e.,
\ba\n{13} &&   \bar X \frac{\partial X}{\partial \bar X} + \bar Y
\frac{ \partial X}{\partial \bar Y} = \frac {X}{n}\n{con1}
\\ && \bar X \frac{\partial Y}{\partial \bar X} + \bar Y
\frac{ \partial Y}{\partial \bar Y} = \frac {Y}{n}. \n{con2}
\ea

For a general massive fermionic field whose potential depends only on $X$, the solution of (\ref{con1}) along with (\ref{trf}) leads to the general FIT
\ba
\n{tx2}
&&\bar X=X^n,\\
\n{tv2}
&&\bar V=n^2V+n^2m\sqrt{X}-\bar m X^{n/2},
%&&\bar m=n^2m \n{tm}.
\ea
where we have normalized to one the multiplicative integration constant in Eq. (\ref{tx2}). Because the fermionic masses represent two different particles, we consider both masses as free parameters of the model. Then, Eqs. (\ref{tx2}) and (\ref{tv2}) define a FIS of the Einstein-Dirac dynamics. Also, from Eqs.  (\ref{r}) and (\ref{pxy}) the equation of conservation (\ref{con}) becomes
\be
\left(m \sqrt{X} + 2 X  {dV\over dX}\right) (\dot X  + 6 X H) = 0.
\ee
From (29), it is easy to see that the two particular cases  (i){$V =V_0$},
$m \ne 0$ and (ii) $m = 0$, {$V \neq V_0$} lead to the same solution, $X \propto a^{-6}$,
so that  $V$ can be treated interchangeably as a function of $X$ or a function of $a$.

%%%%%%%%%%%%%%%%%%%%%%%%%%%%%%%%%%%%%%%%%%%%%%%%%%%%%%%%%
\subsection{Fermionic power law group}
%%%%%%%%%%%%%%%%%%%%%%%%%%%%%%%%%%%%%%%%%%%%%%%%%%%%%%%

Following the same methodology as the bosonic scalar field, let us  investigate  two simple cases, (i) the free massive fermionic particle and (ii) the massless fermionic particle with a polynomial potential.

In the free case, the energy density (\ref{r}) and pressure (\ref{pxy}) of the fermionic particle are given by
\be
\n{v=0}
\rho=m\sqrt{X},  \qquad  p=0, \qquad V=0.
\ee

\no Solving the Einstein equations we obtain
\be
\n{seed2}
a^{(\pm)}=a_0(\pm\,t)^{2/3},  \qquad X=\frac{16}{9m^2}\left(\frac{a_0}{a}\right)^6,
\ee
where $a_0$ is an integration constant. Suppose now that we are interested in comparing this cosmological model with another one driven by a potential, which is assumed to  depend only on $X$. This can be done by inserting the seed configuration (\ref{v=0})-(\ref{seed2})  into the transformations (\ref{tr3}), (\ref{tr}), (\ref{tha}) and (\ref{tx2})-(\ref{tv2}). Then, we find the full set of power law solutions along with the potential, energy density and pressure characterizing the transformed fermionic configuration:
\ben
\n{s2}
&\bar a^{(\pm)}=\bar a_0(\pm\, t)^{2n/3}, \qquad  \bar a_0=a_0^n,\\
&\bar X=\left(\frac{16}{9m^2}\right)^n\left(\frac{\bar a_0}{\bar a}\right)^6,\\
&\bar V=n^2m\bar X^{1/2n}-\bar m\bar X^{1/2},\\
&\bar\rho=n^2m\bar X^{1/2n}, \qquad
\bar p=\left(\frac{1}{n}-1\right)\bar \rho,\n{p2}
\een
while the Einstein equations for the new configuration are
\ben
\n{n00}
3\bar H^2=\bar\rho, \qquad -2\dot{\bar {H}}=\bar\rho+\bar p.\n{n11}
\een
{The transformed potential $\bar V$ has a vanishing limit for large cosmological time matching smoothly with the seed solution (\ref{v=0})-(\ref{seed2}). While the seed solution (\ref{v=0})-(\ref{seed2}) may be identified with some kind of fermionic dust matter the new solution (\ref{s2})-(\ref{p2}) may be associated with a fermionic barotropic fluid whose equation of state is $\bar\ga=1+\bar p/\bar\rho=1/n$.}

The Dirac equations in a FRW background for the seed and transformed configurations read
\be
\n{d}
 \dot\psi+{3\over2}H\psi+\imath m\gamma^0\psi+\imath\gamma^0{dV\over
 d{\overline\psi}}=0,
 \ee
\be
\n{bd}
 \dot{\bar\psi}+{3\over2}\bar H\bar\psi+\imath \bar m \gamma^0\bar\psi+\imath\gamma^0\frac{d\bar V}
 {d{\bar{\ov\psi}}}=0,
 \ee
along with the corresponding equation for the adjoint $\ov\psi$ and $\bar{\ov\psi}$. In order to preserve the characteristics  of the fermionic particles, we assume that the Dirac matrices remain invariant under a FIS transformation.

Solving the Dirac equation (\ref{bd}) for the new configuration (\ref{s2})-(\ref{p2}), we obtain the spinor field solution
\par
\ben
\bar\psi (t) =\left(\frac{\bar a_0}{\bar a}\right)^{3/2}
\left(\begin{array}{c}
  \bar b_{1}\,e^{-\imath nm\tau_n}\\
  \bar b_{2}\,e^{-\imath nm\tau_n}\\
  \bar d^{*}_{1}\,e^{\imath nm\tau_n}\\
  \bar d^{*}_{2}\,e^{\imath nm\tau_n}\\
\end{array}\right),
\n{spinor}
\een
where
\be
\n{+}
|\bar b_1|^2+|\bar b_2|^2-|\bar d_1|^2-|\bar d_2|^2=\left(\frac{4}{3m}\right)^n>0,
\ee
and
\be
\n{tn}
\tau_n=\left(\frac{4}{3m}\right)^{1-n}\int \left(\frac{\bar a_0}{\bar a}\right)^{3(1-n)/n}\,dt.
\ee
\no For $n=1$ the latter becomes the spinor field  corresponding to the seed solution (\ref{v=0})-(\ref{seed2}).

Now, we consider the massless case and investigate a spinor field driven by a potential $V=V(X)$ depending on $X$. Here, the energy density $\rho$ and the pressure $p$ associated with the spinor field are given by:
\be \n{a} \rho  = V, \,\,\,\,\,\,\, p =2 X \frac{d V}{dX} - V. \ee

\noindent The imposition of the energy density transformation
$ \bar \rho = n^2 \rho$ reduces the transformation rules (\ref{tx2})-(\ref{tv2}) to
\be \n{tpi} \bar X=X^n,\qquad \bar V = n^2 V-\bar m X^{n/2}. \ee

By choosing the seed potential $V=V_0 X^\al$, with $V_0$ and $\al$ constants, we obtain
 \be
 \n{m=0}
 \rho=V_0 X^\al,\qquad p=(2\al-1)\rho,
 \ee
 and from the  Einstein equations it follows
 \be
 a^{(\pm)}=a_0(\pm\,t)^{1/3\al},\qquad X={1\over(3\alpha^2V_0)^{1/\alpha}}\left({a_0\over a}\right)^6.
 \ee

In this case power law solutions for the transformed fermionic configuration read
 \ben
\n{s3}
&\bar a=\bar a_0(\pm t)^{n/3\alpha}, \qquad \bar a_0=a_0^n,\\
&\bar X=\left(3\alpha^2V_0\right)^{-n/\alpha}\left(\frac{\bar a_0}{\bar a}\right)^6,\\
%&\bar X=\left(\frac{1}{3\alpha^2V_0}\right)^{n/\alpha}\left(\frac{\bar a_0}{\bar a}\right)^6,\\
&\bar V=n^2 V_0\bar X^{\alpha/n}-\bar m \bar X^{1/2},\\
&\bar\rho=n^2V_0\bar X^{\alpha/n}, \n{p3} \qquad \bar p=\left({2\alpha\over n}-1\right)\bar \rho.
\een

In the massless case the solution of the Dirac equation (\ref{bd}) for the latter  configuration (\ref{s3})-(\ref{p3}) is given by

\par
\ben
\bar\psi (t) =\left(\frac{\bar a_0}{\bar a}\right)^{3/2}
\left(\begin{array}{c}
  \bar b_{1}\,e^{-\imath n\tau_n}\\
  \bar b_{2}\,e^{-\imath n\tau_n}\\
  \bar d^{*}_{1}\,e^{\imath n\tau_n}\\
  \bar d^{*}_{2}\,e^{\imath n\tau_n}\\
\end{array}\right),
\n{spinor2}
\een
where
\be
\n{+}
|\bar b_1|^2+|\bar b_2|^2-|\bar d_1|^2-|\bar d_2|^2=\left(3\al^2 V_0\right)^{-n/2\al}>0,
\ee
and
\be
\n{tn}
\tau_n=2\al V_0\left(3\al^2 V_0\right)^{-1+n/2\al}\int \left(\frac{\bar a_0}{\bar a}\right)^{3(2\al-n)/n}\,dt.
\ee
 In summary, in both the free and massless cases we have found that the Einstein-Dirac equations admit a power law group. In the former it is generated by fermionic dust matter, while in the latter it is generated by a
polynomial potential $V\propto X^\alpha$. In some sense, there is a resemblance between a scalar field driven by an
exponential potential and a massless Dirac field driven by a polynomial potential. This comparison is justified by the fact that both induce equations of state of the form $p\propto\rho$. Due to the importance of the use of power law solutions in cosmological models, it would be interesting to investigate the existence of power law groups for other kinds of fields.

%%%%%%%%%%%%%%%%%%%%%%%%%%%%%%%%%%%%%%%%%%%%%%
\subsection{Cosmological constant group}
%%%%%%%%%%%%%%%%%%%%%%%%%%%%%%%%%%%%%%%%%%%%%%

In this subsection we consider a mixture of the above two examples namely, the fermionic power law group in the free and massless cases summarized in the seeds given by Eqs. (\ref{v=0}) and (\ref{a}). The former leads to a pressureless dust dominated solution whereas the latter leads to a de Sitter solution in the constant potential case. It is well known that the $\La$CDM cosmological model interpolates between these two asymptotic behaviors. Thus, using this cosmological model as a seed one
\ben
\n{L}
&V=\La, \qquad  \rho=m\sqrt{X}+\La, \qquad p=-\Lambda,\\
%\rho=m\sqrt{X}+V=\frac{4}{3}\left(\frac{a_0}{a}\right)^3+\La,
\n{Lxa}
&X=\frac{16}{9m^2}\left(\frac{a_0}{a}\right)^6,\\
&a=a_0\left[\frac{\pm 2}{\sqrt{3\La}}\sinh{\frac{\sqrt{3\La}}{2}\,t}\right]^{2/3},
\een
where $\La$ is the cosmological constant, the FIS gives
\ben
\n{newV}
&\bar V=n^2\La+n^2m\bar X^{1/2n}-\bar m\sqrt{\bar X},\\
&\bar\rho=n^2\left(m \bar X^{1/2n}+\La\right), \\
&\bar p=n(1-n)m\bar X^{1/2n}-n^2\Lambda,\\
&\bar X=\left(\frac{16}{9m^2}\right)^n\left(\frac{\bar a_0}{\bar a}\right)^{6},\\
&\bar a=\bar a_0\left[\frac{\pm 2}{\sqrt{3\La}}\sinh{\frac{\sqrt{3\La}}{2}\,t}\right]^{2n/3},
\een

Asymptotically, in the limit $t\to\infty$, the new potential tends to $\bar V\to n^2\La$ for $n>0$. Therefore, the last set of equations shows that there exists a large set of models  which generalize the   $\La$CDM one, i.e., the class of model generated by the set of FIT (\ref{tr}) with cosmological constant $\bar\La\approx n^2\La$. They describe universes containing a perfect fluid and with cosmological
constant evolving as a power-law dominated phase at early times, where the
scale factor behaves as $\bar a\propto t^{2n/3}$. Finally, such universes
end in a stable de Sitter accelerated expansion scenario with a scale factor
behaving as $\bar a\propto e^{n\sqrt{\La/3}\,t}$, representing a possible
dark energy final stage.

Finally, for the new configuration (\ref{s2})-(\ref{p2}) the spinor field solution of the Dirac's equation (\ref{bd}) becomes
\par
\ben
\bar\psi (t) =\left(\frac{\bar a_0}{\bar a}\right)^{3/2}
\left(\begin{array}{c}
  \bar b_{1}\,e^{-\imath nm\tau_n}\\
  \bar b_{2}\,e^{-\imath nm\tau_n}\\
  \bar d^{*}_{1}\,e^{\imath nm\tau_n}\\
  \bar d^{*}_{2}\,e^{\imath nm\tau_n}\\
\end{array}\right),
\n{spinor3}
\een
where
\be
\n{+}
|\bar b_1|^2+|\bar b_2|^2-|\bar d_1|^2-|\bar d_2|^2=\left(\frac{4}{3m}\right)^n>0,
\ee
and
\be
\n{tn3}
\tau_n=\left(\frac{4}{3m}\right)^{1-n}\int \left(\frac{\bar a_0}{\bar a}\right)^{3(1-n)/n}\,dt.
\ee
This spinor depends on the cosmological constant through the transformed scale factor $\bar a$.

%%%%%%%%%%%%%%%%%%%%%%%%%%%%%%%%%%%%%%%%%%%%%%%%%%%%%%%%%%%%%%%%%%%%%%%%%
\section{Duality and Phantom cosmologies}
%%%%%%%%%%%%%%%%%%%%%%%%%%%%%%%%%%%%%%%%%%%%%%%%%%%%%%%%%%%%%%%%%%%%%%%%

The ``phantomization" process means that the energy density of the expanding cosmological model being analysed must satisfy $\dot\rho>0$, or equivalently the weak energy condition
(WEC) must be violated, so that $(\rho+p)<0$ \cite{Chimento:2003qy}. In terms of the Hubble factor $H$, these two conditions become $H>0$ and
$\dot H>0$.
 Therefore, one can infer two cases, namely, (a) $\rho$ has an asymptote  $\rho\to\Lambda$ for $t\to\infty$, or (b) $\rho$ grows without limit. In the first case  the scale factor tends to a de Sitter solution $a\to \exp{\sqrt{\Lambda/3}\,t}$. In the second case if it is assumed that the energy density has the asymptotic behavior  $\rho\to \rho_0 a^k$ with $k>0$, the asymptotic
 solution of the Friedmann equation reads
 \ben
\n{asin-}
a^{(-)}\to\left[\frac{2\sqrt{3}}{k\sqrt{\rho_0}\,(t_0-t)}\right]^{2/k}, \qquad t<t_0,\\
\n{asin+}
a^{(+)}\to\left[\frac{2\sqrt{3}}{k\sqrt{\rho_0}\,(t-t_0)}\right]^{2/k}, \qquad t>t_0.
\een
 The expanding solution $a^{(-)}$ is defined for $t<t_0$ and ends in a big rip at $t=t_0$, since the scale factor
 diverges at the finite time $t_0$ resulting in a future singularity. By contrast, the contracting solution $a^{(+)}$ begins from a past singularity  at $t=t_0$. To sum up, $a^{(-)}$ is the ``phantomization" of the solution $1/a^{(-)}$ which ends in a big crunch at $t=t_0$. In terms of FIS the ``phantomization" process means that there is a duality  between two solutions of the Friedmann equation, i.e., between $1/a^{(-)}$ and $a^{(-)}$ \cite{Cataldo:2005gb}-\cite{Chimento:2005au}. In fact, the class of FIT generated by $n=-1$:
\be
\n{}
\rho\to\bar\rho=\rho,  \quad  p+\rho\to \bar p+\bar\rho=-(p+\rho),
\ee
with
\be
\n{}
H\to \bar H=-H,\quad a\to \bar a=1/a,
\ee
transforms a contracting scale factor, $H<0$, satisfying the WEC,  $\rho+p>0$ into an expanding one, $\bar H>0$, which violates the WEC, so that $\bar\rho+\bar p<0$. The latter, dubbed the dual transformation, is crucial because of its applicability as a method of transforming a conventional cosmological model into a phantom one by performing a FIT.

In the bosonic case the expressions
(\ref{tp}) and (\ref{tv}) become
\ba
\n{tpph}
&&{\dot {\bar\phi}}\,^2=-\dot {\phi}^2,\\ &&
\n{tvph}
\bar V(\bar
\phi)=\dot\phi^2+V(\phi).
\ea
\no and this implies that  $\bar \phi=i\phi$. That is, the transformed
scalar field is
related to the original one by a Wick rotation, and we finally have
\be
\bar\rho=\frac{1}{2}\dot{\bar\phi}\,^2+{\bar V}(\bar\phi).
\ee The sign in
the kinetic part of the energy density, $\dot{\bar\phi}\,^2=-\dot{\phi}\,^2<0$, indicates that what we actually
have now is a {\it phantom} cosmology, with a phantom field $\bar\phi$, driven by the potential $\bar V(\bar\phi)$, which is a real function of $\phi$.

For the fermionic matter, in the case where the potential depends on $X$, the ``phantomization" induces the following transformation rules for the variables $X$ and $V$
\ben
&&\bar X=\frac{1}{X},\\&&
\bar V=m\sqrt{X}\left(1-\frac{1}{X}\right)+V.
\een
and
\be
\n{ga}
\bar\ga=-\ga,
\ee
for the barotropic index of the spinor field. Hence, there is a duality between a contracting universe filled with an ordinary spinor field with $\ga>0$, and an expanding universe filled with a phantom spinor field with $\bar\ga<0$.

%%%%%%%%%%%%%%%%%%%%%%%%%%%%%%%%%%%%%%%%%%%%%%%%%%%%%%
\section{Conclusion}
%%%%%%%%%%%%%%%%%%%%%%%%%%%%%%%%%%%%%%%%%%%%%%%%%%%%%%

    We have found the symmetry transformations under which the Einstein equations, for a spatially flat FRW space-time filled with  bosonic or  spinor fields, preserve their form. This group of transformations has been used to obtain  power-law solutions from a seed one, for bosonic  or fermionic fluids. We have given special attention to the $n=-1$ case, which expresses the  duality between non-accelerated and accelerated scenarios and vice versa. Thus, starting from a contracting spatially flat FRW cosmological model we get, after using the dual transformation, a super-accelerated  spatially flat FRW cosmological model, i.e., the ``phantomization" of the model. We have shown that bosonic and fermionic fields behave differently under a dual transformation, the former becomes imaginary, whereas the latter changes the sign of its phase and inverts its asymptotic limits. Here we have extended the analysis
of the bosonic case, previously studied in the works \cite{Chimento:2003qy} and \cite{Chimento:2004br}, by discussing the role of the seed fluid, the duality and ``phantomization" of the solutions.
{The analysis of the cosmological constant group with a fermionic seed fluid indicates that one can obtain a model for a universe which evolves from a power-law dominated phase to a stable de Sitter accelerated expansion.}

%%%%%%%%%%%%%%%%%%%%%%%%%%%%%%%%%%%%%%%%%%%%%%%%%%%
\ack
%%%%%%%%%%%%%%%%%%%%%%%%%%%%%%%%%%%%%%%%%%%%%%%%%%%

The authors acknowledge the partial
support under project 24/07 of the  agreement SECYT (Argentina) and CAPES 117/07 (Brazil).
LPC thanks the University of Buenos Aires for partial support under
project X224, and the Consejo Nacional de Investigaciones
Cient\'{\i}ficas y T\'ecnicas under project 5169. FPD and GMK acknowledge the support by
Conselho Nacional de Desenvolvimento Cient\'\i fico e Tecnol\'ogico (CNPq). { The authors acknowledge the comments and suggestions of an anonymous referee.}

\end{document}